\documentclass{article}
\usepackage{spconf,amsmath,graphicx}
\usepackage{multirow}
\usepackage{booktabs}
\usepackage[table,xcdraw]{xcolor}
\usepackage{amsmath,amsthm,amsfonts,amssymb,bm}
\usepackage{hyperref,setspace}
\definecolor{mycolor}{RGB}{153,153,153}
\hypersetup{
colorlinks=true,
}
\usepackage{amssymb}
\DeclareMathOperator*{\argmin}{argmin}
\usepackage{threeparttable}

\usepackage[ruled]{algorithm2e}

\usepackage[
backend=biber,
style=ieee,
maxbibnames=5,
maxcitenames=5,
doi=false,isbn=false,url=false,eprint=false
]{biblatex} 
\defbibheading{bibliography}[\refname]{}
\DeclareSourcemap{
	\maps[datatype=bibtex, overwrite=true]{
		\map{
			\step[fieldsource=booktitle,
			match=\regexp{.*Interspeech.*},
			replace={Proc. Interspeech}]
			\step[fieldsource=journal,
			match=\regexp{.*INTERSPEECH.*},
			replace={Proc. Interspeech}]
			\step[fieldsource=booktitle,
			match=\regexp{.*ICASSP.*},
			replace={Proc. ICASSP}]
			\step[fieldsource=booktitle,
			match=\regexp{.*icassp_inpress.*},
			replace={Proc. ICASSP (in press)}]
			\step[fieldsource=booktitle,
			match=\regexp{.*International.*Conference.*on.*Acoustics,.*Speech.*and.*Signal.*Processing.*},
			replace={Proc. ICASSP}]
			\step[fieldsource=booktitle,
			match=\regexp{.*International.*Conference.*on.*Learning.*Representations.*},
			replace={Proc. ICLR}]
			\step[fieldsource=booktitle,
			match=\regexp{.*International.*Conference.*on.*Machine.*Learning.*},
			replace={Proc. ICML}]
			\step[fieldsource=booktitle,
			match=\regexp{.*International.*conference.*on.*machine.*learning.*},
			replace={Proc. ICML}]
			\step[fieldsource=booktitle,
			match=\regexp{.*Automatic.*Speech.*Recognition.*and.*Understanding.*},
			replace={Proc. ASRU}]
			\step[fieldsource=booktitle,
			match=\regexp{.*Spoken.*Language.*Technology.*},
			replace={Proc. SLT}]
			\step[fieldsource=booktitle,
			match=\regexp{.*Speech.*Synthesis.*Workshop.*},
			replace={Proc. SSW}]
			\step[fieldsource=booktitle,
			match=\regexp{.*workshop.*on.*speech.*synthesis.*},
			replace={Proc. SSW}]
			\step[fieldsource=booktitle,
			match=\regexp{.*Advances.*in.*neural.*information.*processing.*},
			replace={Proc. NeurIPS}]
			\step[fieldsource=booktitle,
			match=\regexp{.*Advances.*in.*Neural.*Information.*Processing.*},
			replace={Proc. NeurIPS}]
			\step[fieldsource=booktitle,
			match=\regexp{.*Workshop.*on.* Applications.* of.* Signal.*Processing.*to.*Audio.*and.*Acoustics.*},
			replace={Proc. WASPAA}]
			\step[fieldsource=publisher,
			match=\regexp{.+},
			replace={{}}]
			\step[fieldsource=month,
			match=\regexp{.+},
			replace={{}}]
			\step[fieldsource=location,
			match=\regexp{.+},
			replace={{}}]
			\step[fieldsource=address,
			match=\regexp{.+},
			replace={{}}]
			\step[fieldsource=organization,
			match=\regexp{.+},
			replace={{}}]
		}
	}
}

\title{Discretization and Re-synthesis: an alternative method to solve the Cocktail Party Problem}
\name{
    Jing Shi$^{1}$,
    Xuankai Chang$^{2}$,
    Tomoki Hayashi$^{3,5}$, 
    Yen-Ju Lu$^{2,4}$,
    Shinji Watanabe$^{2}$,
    Bo Xu$^{1}$}
\address{$^{1}$Institute of Automation, Chinese Academy of Sciences (CASIA),
$^{2}$Carnegie Mellon University, \\
$^{3}$Nagoya University, 
$^{4}$Academia Sinica, 
$^{5}$Human Dataware Lab. Co., Ltd.}
\begin{document}
\ninept
\maketitle

\begin{abstract}

Deep learning based models have significantly improved the performance of speech separation with input mixtures like the cocktail party. Prominent methods (e.g., frequency-domain and time-domain speech separation) usually build regression models to predict the ground-truth speech from the mixture, using the masking-based design and the signal-level loss criterion (e.g., MSE or SI-SNR). 
%We propose here 
This study demonstrates, for the first time, that the synthesis-based approach can also perform well on this problem, with great flexibility and strong potential.
Specifically, we propose a novel speech separation/enhancement model based on the recognition of discrete symbols, and convert the paradigm of the speech separation/enhancement related tasks from regression to classification. By utilizing the synthesis model with the input of discrete symbols, 
%we transfer the target speech of speech separation/enhancement into a sequence of discrete symbols. 
after the prediction of discrete symbol sequence, each target speech could be re-synthesized. 
%Thanks to the use of discrete symbols, we can use various powerful techniques developed in sequence generation tasks (e.g., Speech translation, automatic speech recognition) such as beam search and fusions with a language model. 
%Meanwhile, the training of models become much more efficient compared to the same architecture used in existing speech separation task. 
Evaluation results based on the WSJ0-2mix and VCTK-noisy corpora in various settings show that our proposed method can steadily synthesize the separated speech 
%of good listening quality
with high speech quality
%, nearly 
and without any interference, which is difficult to avoid in regression-based methods. In addition, with negligible loss of listening quality, the speaker conversion of enhanced/separated speech could be easily realized through our method.
%, making our method 

\end{abstract}

\section{Introduction}
\vspace{-0.15cm}
With the popularity of speech-related intelligent devices and related applications, 
%the research of speech front-end technology
front-end processing has become a popular research topic~\cite{SpeechProcessingforDigitalHomeAssistants}. Among them, 
%aiming at solving the cocktail party problem,
%and the representative speech separation/enhancement problem,
a batch of methods based on end-to-end deep learning have emerged to solve the cocktail party problem~\cite{Kolbaek2017Multitalker_pit,hershey2016deepclustering, yu2017permutation,luo2018tasnet,luo2019dual,subramanian2019speech}.
Compared to conventional approaches like computational auditory scene analysis~\cite{bregman1990auditory,bregman1994auditory} and non-negative matrix factorization~\cite{schmidt2006single}, end-to-end models are entirely data-driven, achieving remarkable improvement in speech quality and intelligibility.
%the quality of the separated/enhanced speech.

Although satisfactory results have been observed on laboratory corpus, the design patterns followed by existing prominent methods (e.g., frequency-domain and time-domain speech separation) will also introduce corresponding problems or risks, making them unable to be used reliably in real-world scenarios. Specifically, most existing methods usually build regression models to recover the ground-truth speech as much as possible, with the masking-based design and the signal-level loss criterion, e.g., the mean square error (MSE) between the spectrograms or negative scale-invariant signal-to-noise ratio (SI-SNR) between the waveforms.  Normally, the essence of regression based speech separation/enhancement methods is to map the mixed/noisy speech into a continuous high-dimensional space so that the speech of different source can be better separated. 
However, in a complex auditory scene, the representations of different sources in the mixture speech may be very similar in some segments or even the whole utterance, making it difficult for regression-based methods to exclude the interference from other sources completely.
For this reason, in our preliminary experiments and observation from others, the well-trained SOTA time-domain methods suffer in separating similar speakers' mixed speech and some non-stationary noise.

\begin{figure}
    \centering
    \resizebox{1.0\linewidth}{!}{
    \includegraphics[width=\linewidth]{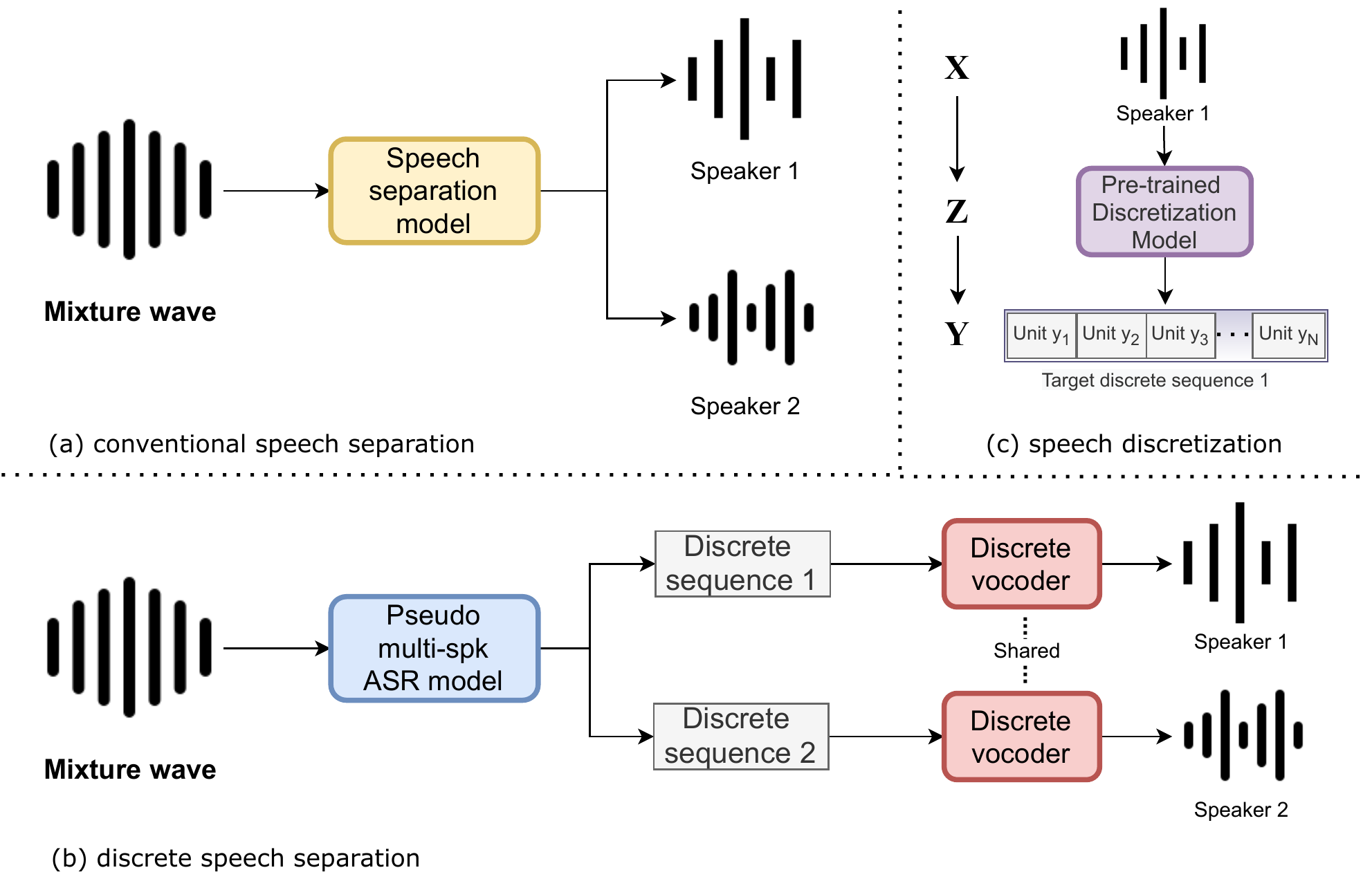}}
    \caption{Illustration of conventional speech separation model and our proposed discretization-synthesis model.}
    \label{fig:model}
\vspace{-0.5cm}
\end{figure}

%\vspace{-0.1cm}
To tackle these agonizing challenges, one possible solution is to replace the direct signal-level masking over the original input.
In this work, we propose a novel speech separation/enhancement model based on the recognition of discrete symbols, converting the paradigm of the speech frontend related tasks from regression to classification. Without loss of generality, we show the case with two speakers condition of our proposed method in Fig.\ref{fig:model}.  Specifically, by utilizing the great reconstruction ability of the vector quantized variational autoencoder (VQ-VAE)~\cite{vq-vae} or HuBERT\cite{hsu2021hubert} + HiFi-GAN\cite{kong2020hifigan} model, we transfer the target speech of speech separation model into a sequence of discrete symbols. With the estimation of discrete symbol sequence, each target speech could be re-synthesized with optional transferred  style. 
%Thanks to the use of discrete symbols, 
%we can use various powerful techniques developed in sequence generation tasks (e.g., Speech translation, automatic speech recognition) such as beam search and fusions with a language model. Meanwhile, 
%the training of models become much more efficient compared to the same architecture used in existing speech separation task. 

The experimental evaluation with the WSJ0-2mix~\cite{isik2016deepclustering} and VCTK-noisy~\cite{ValentiniBotinhao2017vctknoisy} corpus in various settings shows that the proposed method could steadily synthesize the separated speech of good listening quality, nearly without any interference, which is difficult to avoid in masking based methods. In addition, with almost no impact on perception and cognition evaluation, the speaker conversion of separated speech could be easily realized through our model.

The core contribution of this work is to articulate an alternative direction, besides the dominating masking-based regression paradigm, with synthesis after discretization for speech enhancement/separation task. We also demonstrate the benefit it gets, along with the difficulty it faces with current mainstream evaluation metrics. Our proposed method presents a promising avenue for exploring solutions to the problem. The sound demo for this paper is available at \href{https://shincling.github.io/discreteSeparation/}{https://shincling.github.io/discreteSeparation/}.

\vspace{-0.4cm}
\section{methodology}
\label{ssec:method}

\vspace{-0.2cm}
\subsection{Background Methods}
\vspace{-0.15cm}
\subsubsection{Discretization of Speech}\label{sssec:discret}
\vspace{-0.15cm}
Similar to the low-bitrate speech codecs~\cite{atal1971speechlowbite1,mccree1996lowbit2,kleijn2018wavenetlowbit3,lim2020robustlowbit4}, speech discretization aims at encode the speech input into a discrete sequence, shown in Fig.\ref{fig:model}(c). %Commonly, we want to encode as much information as possible, and ideally be able to restore the original speech signal.
Ideally, we want to encode as much information as possible so that the original speech can be restored.
In this paper, we use two methods to gets the discrete units from raw speech: VQ-VAE~\cite{vq-vae} and HuBERT~\cite{hsu2021hubert}.

\textit{VQ-VAE} model can convert a sequential input with an arbitrary length into a downsampled sequence of discrete symbols, and precisely reconstruct the input from the downsampled discrete symbol sequence~\cite{vq-vae} in a unified structure. \textit{HuBERT} is a self-supervised learning based model trained with masked continuous audio signals. It shows superior performance across multiple tasks such as ASR, spoken language modeling, and speech synthesis~\cite{yang2021superb}. For the HuBERT model, we use the k-means algorithm to cluster the extracted latent representations 
into the clustering center while the VQ-VAE directly use the nearest centroid in VQ-codebook. 

\vspace{-0.25cm}
\subsubsection{Discrete Vocoder}\label{sssec:vocoder}
\vspace{-0.15cm}
As illustrated in Fig.\ref{fig:model}(b), the synthesis of the speech from discrete symbols can be formed as the traditional text-to-speech (TTS) models, which usually produce Mel-spectrogram autoregressively given textual features as input. Next, a vocoder is applied to reconstruct the phase from the Mel-spectrogram. %(e.g., Griffin-Lim~\cite{griffin1984signal_stft}, WaveNet~\cite{oord2016wavenet_raw}, WaveGlow~\cite{prenger2019waveglow}, or HiFi-GAN~\cite{kong2020hifigan}).
In this study, we propose to use the learned discrete speech units as an input to a vocoder module with no spectrogram estimation. The discrete symbols are extracted with the pre-trained self-supervised models, e.g., VQ-VAE or HuBERT model, and we use MelGAN~\cite{kumar2019melgan} and HiFi-GAN~\cite{kong2020hifigan} as architecture for the vocoder module accordingly. 
It is worth mentioning that, to synthesize the original speech as much as possible with various speakers, the speaker identities information can be utilized as the condition in Vocoder.

\vspace{-0.25cm}
\subsection{Proposed Discrete Separation } \label{ssec:discretess}
\vspace{-0.15cm}
Our proposed method shown in Fig.\ref{fig:model}(b) can be divided into two modules: discrete sequence prediction and speech re-synthesis\footnote{Without loss of generality, we set the output number of sources to 2 in this figure, which can be single or more speakers.}. Assume the mixed/noisy signal is $\mathbf{M}$, and the target clean signal to be $\mathbf{X}^k$, where $k$ is the index of the target speaker. 

\vspace{0.1cm}
\noindent\textbf{Discrete sequence prediction} 
Before training, as illustrated in Fig.\ref{fig:model}(c), we first build the tool that can discretize continuous waveform $\mathbf{X}$ into the sequence of the discrete symbols $\mathbf{Y}=\{\mathbf{y}_1, ...,\mathbf{y}_n, ..., \mathbf{y}_N\}$, i.e., centroid IDs in the VQ-codebook or HuBERT clustering
%, ideally without much informative loss
.  As mentioned in Sec.\ref{sssec:discret}, we use either our own-trained VQ-VAE encoder or the pre-trained HuBERT model to sever as the \textit{discretizer} here. This process could be formulated as follows: \vspace{-0.4cm}
\begin{align}
\mathbf{Z} &= \{{\mathbf{z}_1,..., \mathbf{z}_n,..., \mathbf{z}_N}\}= F(\mathbf{X}), \\
    \mathbf{y}_n &= Q(\mathbf{z}_n) = i 
    %\ \  \mathrm{where} \ \ i 
    = \argmin_{j} \|\mathbf{z}_n - \mathbf{e}_j \|,    
\end{align}
where $F$ is the feature extraction function (VQ-VAE encoder or HuBERT), $Q$ the discretization function, $\mathbf{e}_j$ the $j$-th centroid in codebook or clustering, and $N$ the total length of discrete sequence $\mathbf{Y}$, which is down-sampled from the original length $\mathbf{T}$. To summarize, this  here acts as: $\mathbf{X}\longrightarrow\mathbf{Z} \longrightarrow \mathbf{Y}$.
With the \textit{discretizer}, in the training phase, the ground-truth speech $\mathbf{X}$ of each speaker could be transferred into a target discrete sequence $\mathbf{Y}$. 
%In practice, we found both of them can reconstruct the original input waveform with a very high quality, through the \textit{Vocoder} described in Sec.\ref{sssec:vocoder}. 

%to  train the non-autoregressive GAN-based VQ-VAE model using clean speech waveforms in the mixture. The aim of the VQ-VAE model is to build a tool that can discretize continuous waveform input without much informative loss, which could be used directly by the following downstream model.  Besides that,  this discretized sequence can almost perfectly reconstruct the speech of original speech. After training the VQ model, the VQ-encoder could convert all speech reference waveforms into sequences of the discrete symbols, i.e., centroid IDs in the VQ-codebook.   
With these discrete symbols $\mathbf{Y}$ for each speaker as the target rather than the original continuous clean speech $\mathbf{X}$, the speech separation/enhancement task could be converted into the task of multi/single-speaker speech recognition, which we use the pseudo ASR module in Fig.~\ref{fig:model}(b) to conduct.  
For the module of pseudo ASR, taking the input $\mathbf{M}$ from the noisy/mixed speech, this module acts to estimate the posterior probability of the discrete sequence as follows:
\vspace{-0.6cm}
\begin{align}
p(\mathbf{\hat{Y}|M}) =& \prod_{n=1}^{N}p(\mathbf{\hat{Y}}_n|\mathbf{M}) \ \ \mathrm{where} \ \  \mathbf{\hat{Y}} = \{\hat{\mathbf{y}}_1,...,\hat{\mathbf{y}}_N\}.
\end{align}
Thanks to the \textit{discretizer}'s convolutional architecture, the alignment between the target sequence $\mathbf{Y}$ and the corresponding mixture/noisy speech $\mathbf{M}$ is guaranteed. 
Based on this, we adopt the pseudo ASR module which performs as the frame-by-frame processing similar to hybrid HMM/DNN ASR framework.
To be specifc, for pseudo ASR module, our framework consisting of \textit{Encoder}, \textit{Separator}\footnote{To keep it simple, we use the term \textit{Separator} herein for both speech separation and speech enhancement task. } and \textit{Classifier}. On the contrast to the conventional mainstream speech separation framework with \textit{Encoder} (STFT/Conv Encoder), \textit{Separator} and \textit{Decoder} (iSTFT/Conv-Decoder), our proposed framework replaces the \textit{Decoder} with \textit{Classifier}. This is because we do not need to reconstruct the signal at this module, but to predict the discrete symbols. 
In addition, as mentioned before, the speaker information can also be used in \textit{Vocoder} to reconstruct the speech.  Based on that, the speaker prediction is also implemented together within this module by averaging the latent features after the \textit{Separator}. 
\vspace{0.1cm}

\noindent\textbf{Speech re-synthesis} 
After the prediction of discrete symbols (also the optional speaker), the discrete vocoder described in Sec.\ref{sssec:vocoder} is used to synthesize the final predicted speech. In practice, the vocoder could be highly flexible with different training settings. For instance, if we use the homologous data used in the separation part, the vocoder could reconstruct speech as closely as possible to the original data set. By contrast, the vocoder could also be trained with heterogeneous data given the unified discrete symbols, e.g., from the HuBERT, resulting in the re-synthetic speech with the transferred style of the heterogeneous data. In our experiments, we implemented both the multi-speaker homologous vocoder and a single-speaker heterogeneous vocoder to valid the speciality of speaker-transfer. Besides this simple setting, more styles or types of reconstruction could be explored further, e.g., accent-transfer, tone-transfer, emotion-transfer and so on, which we defer to future work. 

The detailed procedure of our whole proposed framework could been seen in Procedure~\ref{alg}.

\noindent\textbf{2-stage refining}
As we can imagine and through our practice, even though it is almost indistinguishable from hearing, synthesis-based method gets difficulties to accurately recover the original speech at the signal level. To show that our reconstructed speech already contains useful information of the target speech as we hear, here we propose a 2-stage refining strategy. In brief, we use the estimated speech $\mathbf{\hat{X}}$ as the condition (target speech) to further trained a separation (target-speech extraction) model. With the concatenation of mixture signal $\mathbf{M}$ and $\mathbf{\hat{X}}$ as input, we expect the 2-stage model can refine the estimated speech towards the ground-truth signal $\mathbf{X}$. 

\vspace{-0.45cm}
\section{Related Work}
\vspace{-0.3cm}
Our method shares some similarities with exemplar-based speech enhancement~\cite{baby2015exemplar1,baby2015coupledexemplar2,baby2014coupledexemplar3}, which often uses linear combination of the trained speech/noisy combined dictionaries to reconstruct the raw waveform and then to perform enhancement.
To some extent, our discretized sequence can be regarded as an extreme sparse case of linear coefficients, and the Vocoder is much more complex with a GAN-based neural networks rather than the simple summation of the spectrograms or waveforms.    
Recently, the discrete units based methods have been applied for speech synthesis~\cite{hayashi2020discretalk,polyak2021discretedisentangled} and speech translation~\cite{lee2021direct}. Although with a similar discretization and synthesis pipeline, our work focuses more on the extraction of speech-like audios given the complex noisy/mixed speech, rather than the reconstruction or speaker conversion of the clean speech. In particular, we first came up with the multi-source re-synthesis from one single speech (fully-overlapped speech separation task), which is quite different with above works. 

\vspace{-0.4cm}
\section{Experiments}
\vspace{-0.25cm}
Our experiments include two settings: speech separation on WSJ0-2mix corpus and speech enhancement on VCTK-noisy corpus. Both subjective evaluation and objective evaluation are conducted to evaluate the proposed methods.
\vspace{-0.35cm}
\subsection{Implementation details}\label{ssub:imple}
\vspace{-0.1cm}
\subsubsection{Pseudo ASR module}\label{sssub:asr}
\vspace{-0.1cm}
Based on the similarity with closely related speech separation task, two types of architectures have been investigated. 
First, the transformer-based ASR module, similar with the multi-speaker ASR task and the architecture proposed by \cite{karita2019comparative} with FBank base feature extraction and multiple transformer encoder as our \textit{Separator}. Differently, for the prediction of discrete sequence, thanks to the alignment of speech and discrete units, the transformer-based autoregressive decoder is not used, but we just use the encoded feature to predict the probability distribution at each frame with the common cross-entropy loss function. 
Second, the conventional module of speech separation could be directly used in our \textit{Encoder} and \textit{Separator} models. In our experiments, we adopt the same architecture of the DPRNN model \cite{luo2019dual} to better compare our proposed method and existing ones.
Moreover, for the speech separation task, utterance-level PIT \cite{Kolbaek2017Multitalker_pit} is used to determine the best permutation of the output streams predicted by the \textit{Separator+Classifier} . 

In the following section, after trying and comparing different models towards different tasks (separation/enhancement), we choose transformer-based model in speech enhancement experiments with single-speaker and Conv-DPRNN based 2-speaker model in speech separation experiments, which performs better in each task. The details about the architecture could be been in Appendix~\ref{apx:asr}

\vspace{-0.25cm}
\subsubsection{Discrete Vocoder}
\vspace{-0.2cm}

\begin{table}[b]
\vspace{-0.7cm}
\caption{Training settings for the discrete vocoder.}
\label{tab:vocoder}
\resizebox{0.47\textwidth}{!}{%
\begin{tabular}{c|c|c|c|c|c}
\toprule
\textbf{\begin{tabular}[c]{@{}c@{}}Vocoder \\ index\end{tabular}} & \textbf{\begin{tabular}[c]{@{}c@{}}Setting \\ name\end{tabular}} & \textbf{\begin{tabular}[c]{@{}c@{}}Vocoder \\ Architecture\end{tabular}} & \textbf{Training Set} & \textbf{\begin{tabular}[c]{@{}c@{}}Num of \\ Speakers\end{tabular}} & \begin{tabular}[c]{@{}c@{}}Downsample \\ rate\end{tabular} \\ \midrule
\textcolor{blue}{\textcircled{1}} & VQ-VAE & MelGAN & WSJ0 (si\_tr\_s) 8k & 101 & 64 \\ \hline
\textcolor{blue}{\textcircled{2}} & \multirow{4}{*}{Hubert + HiFi-GAN} & \multirow{4}{*}{HiFi-GAN} & WSJ0 (si\_tr\_s) 8k & 101 & 160 \\ \cline{1-1} \cline{4-6} 
\color{blue}{\textcircled{3}} &  &  & VCTK$^*$ 16k & 90 & 160 \\
\cline{1-1} \cline{4-6} \textcolor{blue}{\textcircled{4}} &  &  & LJSpeech 8k & 1 & 160 \\
%\cline{1-1} \cline{4-6} \textcolor{blue}{\textcircled{5}} &  &  & LJSpeech 16k & 1 & 160 \\ 
\bottomrule
\end{tabular}%
}
\begin{tablenotes}
\footnotesize
\item[] $^*$We use the subset of VCTK corpus, with 90 speakers in total.
\end{tablenotes}
\vspace{-0.6cm}
\end{table}
As mentioned before, two types of \textit{Vocoder} setting is tried. In preliminary experiments, VQ-VAE was used with the same architecture proposed by~\cite{hayashi2020discretalk}, and got reasonable results in our task. Then, inspired by the impressive performance with HuBERT model~\cite{yang2021superb} in various speech tasks and the great speech quality synthetized with HiFi-GAN vocoder, we use the combination of HuBERT\&HiFi-GAN architecture in the following experiments. To be specific, we use the HuBERT Large model trained on Libri-Light~\cite{kahn2020librilight} 60k hour without any downstream finetuning to extract the features. Then, we trained a kMeans model with 100 clusters, which will be used to generate the discrete symbols for the corresponding data. The details of different settings in Vocoder is shown in Table~\ref{tab:vocoder}. It is worthy mentioning that although the default pre-trained HuBERT model is only used for the speech with sampling rate of 16kHz, the corresponding extracted discrete symbols $\mathbf{Y}$ could be successfully used to train the vocoder with different sampling rate for the output waveform. From the observation of our trained vocoder, the reconstructed speech still gets very impressive listening quality. All the training speakers in \textit{Vocoder} has \textbf{no-overlap} with the speakers in test for all experiments.

\vspace{-0.35cm}
\subsection{Evaluation Metrics}
\vspace{-0.15cm}
Due to the essence of synthesis-based model, the proposed model is hard to compete against the regression-based methods in the metrics of the signal-level similarity, e.g., SDR, SNR, SI-SNR, STOI and so on. This is because these above metrics are all designed to evaluate how accurate the reconstruction of the original sound signal is, while our synthesized/reconstructed speech is more like the perceptual approximation of human hearing.
Based on these facts, in the following sections, along with the conventional objective evaluation metrics for speech enhancement/separation, we also conduct the subjective evaluation results. Through the evaluation of both sides, we hope to better analyze the advantages and difficulty of the synthesis methods we propose. Besides the STOI, SAR, SDR and SIR, we use an additional open-source MOSNet~\cite{2019mosnet} results to provide some rough simulation of subjective evaluation about the speech quality. More details about the definition of various metrics used in the following sections could be seen in Appendix~\ref{apx:metrics}.

\vspace{-0.35cm}
\subsection{Speech separation results}\label{ssec:exp_ss}
\vspace{-0.15cm}
\begin{table}[]
\caption{The overall objective evaluation in separation performance with our proposed method and some baseline models in WSJ0-2mix validation and test set.}
\label{tab:ss_all}
\resizebox{0.5\textwidth}{!}{%
\begin{tabular}{l|c|c|c|c|c|c}
\toprule
\multicolumn{1}{c|}{\multirow{2}{*}{\textbf{Model}}} & \multirow{2}{*}{\textbf{\begin{tabular}[c]{@{}c@{}}Discrete\\ Vocoder\end{tabular}}} & \multicolumn{5}{c}{\textbf{Separation Performance (cv/tt)}} \\ \cline{3-7} 
\multicolumn{1}{c|}{} &  & \textbf{STOI} & \textbf{SAR} & \textbf{SDR} & \textbf{SIR} & \textbf{MOSNet} \\ \toprule
(1) GT wave & x & - & - & - & - & 3.129/3.216 \\ \hline
\begin{tabular}[c]{@{}l@{}}(2) Baseline: \\ Conv-TasNet\cite{luo2019conv}\end{tabular} & x & 0.937/0.938 & 15.45/14.75 & 14.68/13.78 & 24.35/23.15 &  2.917/3.019\\ \hline
\begin{tabular}[c]{@{}l@{}}(3) Baseline: \\ Conv-DPRNN\cite{luo2019dual}\end{tabular} & x & 0.950/0.960 & 17.41/17.09 & 16.83/16.46 & 27.43/26.90 & 2.953/3.047 \\ \hline
\multirow{2}{*}{\begin{tabular}[c]{@{}l@{}}(4) GT HuBERT \\ discrete  symbols\end{tabular}} & \begin{tabular}[c]{@{}c@{}}Multi-spk \\ WSJ0\textcolor{blue}{\textcircled{2}}\end{tabular} & 0.80/N.A & -10.53/N.A & -11.19/N.A & 9.19/N.A & 2.936/N.A \\ \cline{2-7} 
 & \begin{tabular}[c]{@{}c@{}}Single-spk \\ LJspeech\textcolor{blue}{\textcircled{4}}\end{tabular} & 0.62/0.62& -13.68/-13.52  & -14.94/-14.79  & 5.79/5.78  & 2.982/2.977 \\ \midrule\midrule
\multirow{2}{*}{\begin{tabular}[c]{@{}l@{}}(5) Our-base\\ (Conv-DPRNN)\end{tabular}} & \begin{tabular}[c]{@{}c@{}}Multi-spk \\ WSJ0\textcolor{blue}{\textcircled{2}}\end{tabular} & 0.70/0.67 & -11.95/-12.44 & -12.94/-13.60 & 7.81/6.41 & 2.937/2.976 \\ \cline{2-7} 
 & \begin{tabular}[c]{@{}c@{}}Single-spk \\ LJspeech\textcolor{blue}{\textcircled{4}}\end{tabular} & 0.58/0.58 & -13.67/-13.53  & -15.08/-14.97  & 5.23/5.16  &2.976/2.978  \\ \hline
\multicolumn{1}{r|}{\begin{tabular}[c]{@{}r@{}}+ 2stage \\ Conv-TasNet\end{tabular}} & \begin{tabular}[c]{@{}c@{}}Multi-spk \\ WSJ0 \textcolor{blue}{\textcircled{2}}\end{tabular} & 0.944/0.946 & 16.77/15.97 & 16.17/15.31 & 26.80/25.76 &  3.021/3.119\\ \hline
\multicolumn{1}{r|}{\begin{tabular}[c]{@{}r@{}}+ 2stage \\ Conv-DPRNN\end{tabular}} & \begin{tabular}[c]{@{}c@{}}Multi-spk \\ WSJ0\textcolor{blue}{\textcircled{2}}\end{tabular} & 0.948/0.959 & 17.59/17.13 & 16.98/16.48 & 27.69/27.12 & 3.009/3.092  \\ \bottomrule
\end{tabular}%
}\vspace{-0.5cm}
\end{table}

\begin{table}[!]
\centering
\caption{Separation MOS with different gender combinations for 2-speaker's mixtures from test set of WSJ0-2mix.}
\label{tab:ss_sub_eval}
\resizebox{0.4\textwidth}{!}{%
\begin{tabular}{c|c|c|c}
\toprule
\multirow{2}{*}{\textbf{Methods}}                               & \multicolumn{3}{c}{\textbf{Gender Combination with 2 spkrs}} \\ \cline{2-4} 
             & \textbf{F+M} & \textbf{M+M} & \textbf{F+F} \\ \hline
Ground-Truth & 4.33 ± 0.13  & 4.46 ± 0.12  & 4.27 ± 0.13  \\ \midrule\midrule
Conv-TasNet\cite{luo2019conv}  & 3.45 ± 0.17  & 3.37 ± 0.17  & 3.27 ± 0.18  \\ \hline
Conv-DPRNN\cite{luo2019dual}   & 3.85 ± 0.14  & 3.58 ± 0.16  & 3.53 ± 0.16  \\ \hline
\begin{tabular}[c]{@{}c@{}}Our-base (Conv-DPRNN)\end{tabular} & \textbf{4.12 ± 0.13}         & \textbf{4.22 ± 0.12}        & \textbf{4.10 ± 0.14}        \\ \bottomrule
\end{tabular}%
}
\vspace{-0.35cm}
\end{table}  
For speech separation experiments, we use the WSJ0-2mix corpus, sampled at 8 kHz with the max-length mode, by which we want to maintain the semantic integrity of each target speech. 

The overall objective evaluation is listed in the Table~\ref{tab:ss_all}. The oracle and baseline models are listed as: (1) ground-truth reference waveform; (2) the Conv-TasNet~\cite{luo2019conv}; (3) the Conv-DPRNN~\cite{luo2019dual}; (4) the results with ground-truth Hubert discrete symbols with our trained Vocoder. 

For our models, we set the base model (5) of our proposed methods with the same architecture of Conv-DPRNN~\cite{luo2019dual} in ASR module, while replacing the Conv-Decoder to classifier as mentioned in the Sec.\ref{sssub:asr}. In all our experiments trained for WSJ0-2mix, we found the time-domain based separation models, e.g., Conv-TasNet, DPRNN, Dual-path Transformer, perform better than the transformer-based multi-spk ASR model by an obvious margin. Moreover, the DPRNN performs more stable than the others two models. Based on these results, we use the Conv-DPRNN model as our base model in the separation experiments. 
\begin{table}[]
\centering
\caption{The overlap-ratio of different methods with two thresholds. The lower the threshold, the more sensitive it is to detect overlaps.  }
\label{tab:ss_ovl}
\resizebox{0.48\textwidth}{!}{%
\begin{tabular}{l|c|c|c|c|c|c}
\toprule
\rowcolor[HTML]{FFFFFF} 
{\color[HTML]{FFFFFF} \textbf{Threshold=0.3}} &
  \multicolumn{3}{c|}{\cellcolor[HTML]{FFFFFF}\textbf{valid set ({\color[HTML]{CB0000}Threshold = 0.3 / 0.5})}} &
  \multicolumn{3}{c}{\cellcolor[HTML]{FFFFFF}\textbf{test set ({\color[HTML]{CB0000} Threshold = 0.3 / 0.5})}} \\ \midrule
\textbf{Model} &
  \textbf{\begin{tabular}[c]{@{}c@{}}Speech \\ length(s)\end{tabular}} &
  \textbf{\begin{tabular}[c]{@{}c@{}}Overlap \\ length (s)\end{tabular}} &
  \textbf{Ratio(\%)} &
  \textbf{\begin{tabular}[c]{@{}c@{}}Speech \\ length(s)\end{tabular}} &
  \textbf{\begin{tabular}[c]{@{}c@{}}Overlap \\ length (s)\end{tabular}} &
  \textbf{Ratio(\%)} \\ \midrule\midrule
Conv-TasNet &
  \cellcolor[HTML]{FFFFFF}7.69 &
  0.369 / 0.233 &
  4.80 / 3.04 &
  \cellcolor[HTML]{FFFFFF}7.46 &
  \cellcolor[HTML]{FFFFFF}0.4441 / 0.2566 &
  \cellcolor[HTML]{FFFFFF}5.95 / 3.44 \\ \hline
Conv-DPRNN &
  7.61 &
  0.175 / 0.115 &
  2.30 / 1.51 &
  \cellcolor[HTML]{FFFFFF}7.35 &
  \cellcolor[HTML]{FFFFFF}0.2405 / 0.1594 &
  \cellcolor[HTML]{FFFFFF}3.27 / 2.17 \\ \hline
Our-base &
  \cellcolor[HTML]{FFFFFF}6.75 &
  \cellcolor[HTML]{FFFFFF}\textbf{0.0404 / 0.0102} &
  \cellcolor[HTML]{FFFFFF}\textbf{0.598 / 0.150} &
  \cellcolor[HTML]{FFFFFF}6.60 &
  \cellcolor[HTML]{FFFFFF}\textbf{0.04547 / 0.01067} &
  \cellcolor[HTML]{FFFFFF}\textbf{0.689 / 0.162 } \\ \bottomrule
\end{tabular}%
}
\vspace{-0.5cm}
\end{table}

Unsurprisingly, for the signal-level separation metrics, our synthesis based models shows quite poor performance. We consider there are several reasons: (1) The speech reconstructed by the vocoder, trained with the GAN-based loss criteria, is not fully designed to reconstruct the original waveform but to make it difficult to distinguish with the distribution of real speech; (2) the Vocoder is performed with the condition of predicted known speaker in \textcolor{blue}{\textcircled{2}} and the one external speaker in \textcolor{blue}{\textcircled{4}} from the LJSpeech. These outputted speech with this setting actually could be seen as the audio after speaker conversion, which will inevitably change the original signal; (3) The reconstructed speech gets frame-shift after the vocoder. After the 2-stage refining, the better results compared with the same baseline separation model ((2) and (3)) indicate that the reconstructed speech $\mathbf{\hat{X}}$ from our base model could provide informative clue in signal-level, which is consistent with our hearing's feeling.  

Although the separation metrics from the base model is poor, from the generated samples with either the oracle HuBERT symbols or our predicted HuBERT symbols, we found the speech is quite audible and reasonable. In particular, the re-synthesize speech contains almost no interference, no matter how close the two speakers are in the mixture, which is difficult for the masking-based models to tackle. To better evaluate our discretization-synthesis methods, then we use the subjective evaluation of separation mean opinion score(sMOS), through which we want to show the human judge about the separation degree. In detail, we randomly sampled 20 mixture samples from WSJ0-2mix test set with each gender combinations: 2 females, 2 males and 1 male 1 female. The number of subjects is 23, and that of evaluation samples per each subject is 120 (= 10 samples (from 20 in total) × 3 gender combination × 4 models). Each subject rated the amount of interference of each sample on a 5-point scale: 5 for completely no interference, 4 for almost not interference, 3 for a small amount of interference could be head, 2 for obvious interference could be heard, and 1 for too much interference. We instructed subjects to work in a quiet room and use headphones. We used the WebMUSHRA to build the online tools. The final results of sMOS are listed in Table~\ref{tab:ss_sub_eval}. From the separation MOS, we could find that our proposed method get obvious advantage.  Moreover, our method shows a similar tendency with the Ground-truth results, which indicates the stability across the different gender combination, while baselines are all suffered with the same-gender setting compared with the different gender. 

To further evaluate the purity of our generated samples, we use the overlapped speech length ratio calculated by the Pyannote Overlapped detection tool \cite{bredin2020pyannote}. Lower overlap-ratio means less part of the predicted speech gets over one speaker, indicating the more clean speech. The results shown in Table~\ref{tab:ss_ovl} on all the samples on testset of WSJ0-2mix is also quite promising with nearly an order of magnitude lower than the baselines. For better demonstration of our proposed system, we recommend the readers to visit more samples via: \href{https://shincling.github.io/discreteSeparation/}{https://shincling.github.io/discreteSeparation/}.

\vspace{-0.35cm}
\subsection{Speech enhancement results}
\vspace{-0.15cm}
\label{ssec:enh_exp}
Next, we carried out the speech enhancement experiment on the VCTK\_noisy corpus \cite{ValentiniBotinhao2017vctknoisy}, trained with HuBERT discretizer sing the training set (tr\_26spk). To predict the discrete symbols on the noisy data, we build our single-speaker pseudo ASR model with a 12-layer Transformer Encoder using similar architecture as in end-to-end ASR models \cite{karita2019comparative}. The HiFi-GAN vocoder used was trained on partial VCTK data. We first show the objective evaluation metrics in Table~\ref{tab:enh_objective}. Then we show the noise mean opinion scores (nMOS) collected from 21 testees in Table~\ref{tab:enh_mos}. At last, we show the speech recognition performance of synthesized audios in Table~\ref{tab:enh_asr}. In consistent with separation task, the objective evaluation of our synthesized audio is relatively low. However, the noise MOS are promising, even close to the clean ground truth audios. In terms of intelligibility, We  trained an ASR model using the clean VCTK data. The WER of synthesized audio from predicted discrete symbol sequences are better than the original noisy audio after finetuning ASR model on the generated speech in training set. Please refer to the Appendix \ref{apx:metrics} for the definition of the used metrics. 
\vspace{-0.3cm}
\begin{table}[htp]
    \centering
    \caption{The objective evaluation performance on the cv\_2spk \& tt\_2spk sets of VCTK\_noisy corpus. Four metrics are used: PESQ, scale of signal distortion (SIG), scale of background noise (BAK) and overall effect using scale of MOS (OVL).}
    \resizebox{0.9\linewidth}{!}{\begin{tabular}{c|c|c|c|c}
        \toprule
        Model & PESQ & SIG & BAK & OVL \\
        \midrule
        Conv-TasNet \cite{luo2019conv} & 1.57 / 2.84 & 2.56 / 2.33 & 2.08 / 2.62 & 2.01 / 2.51 \\
        MetricGAN \cite{fu2021metricgan+} & 2.79 / 3.13 & 3.42 / 4.02 & 2.81 / 3.16 & 3.07 / 3.57 \\
        MTL-MIMIC \cite{bagchi2018spectral} & 2.49 / 2.87 & 3.06 / 3.66 & 3.09 / 3.42 & 2.76 / 3.26 \\
        \midrule
        Proposed & 1.26 / 1.21 & 2.81 / 2.70 & 1.71 / 1.62 & 1.97 / 1.87 \\
        \bottomrule
    \end{tabular}
    }
    \label{tab:enh_objective}
\vspace{-0.5cm}
\end{table}

\begin{minipage}{0.5\textwidth}
        \begin{minipage}[t]{0.4\textwidth}
           \centering
           \vspace{0.1cm} \makeatletter\def\@captype{table}\makeatother\caption{Noise MOS for vctk\_noisy corpus.}
            \resizebox{1.1\linewidth}{!}{
            \begin{tabular}{c|c}
        \toprule
        Model & MOS \\
        \midrule
        Clean GT & 4.03 ± 0.13 \\
        Noisy GT & 2.65 ± 0.14 \\
        \midrule
        Conv-TasNet\cite{luo2019conv} & 2.49 ± 0.14 \\
        MetricGAN\cite{fu2021metricgan+} & 3.30 ± 0.14 \\
        MTL-MIMIC\cite{bagchi2018spectral} & 3.39 ± 0.15 \\
        \midrule
        Proposed \\(VCTK vocoder \textcolor{blue}{\textcircled{3}}) & \textbf{4.02 ± 0.14} \\
        \bottomrule
    \end{tabular}\label{tab:enh_mos}
    }
    \end{minipage}
    \hspace{0.05\textwidth}
\begin{minipage}[t]{0.4\textwidth}
\centering
\makeatletter\def\@captype{table}\makeatother\caption{Performance of speech recognition using VCTK-ASR model before / after finetuning on generated speech.}
\resizebox{1.1\linewidth}{!}{
\begin{tabular}{c|c|c}
\toprule
 & \begin{tabular}[c]{@{}c@{}}ASR w/o\\ finetuning\end{tabular} & \begin{tabular}[c]{@{}c@{}}ASR w/\\ finetuning\end{tabular} \\ \midrule
Clean     & 5.8  & 1.4  \\ \hline
Noisy     & 19.5 & 14.6 \\ \hline
GT-symbols        & 14.9 & 8.5  \\ \hline
Predicted & 22.4 & 13.0 \\ \bottomrule
\end{tabular}
\label{tab:enh_asr}}
\end{minipage}
\end{minipage}

\vspace{-0.15cm}
\section{Conclusion and Future Work}
\vspace{-0.25cm}
In this work, we have proposed a discretization-resynthesis framework for speech enhancement and speech separation. %to separate/enhance mixed/noisy speech.
With the predicted discrete sequence for each target speaker, we convert the existing regression-based speech separation/enhancement paradigm into a classification-based one. %, which brings some notable changes accordingly.
Evaluation results on the WSJ0-2mix and VCTK-noisy corpus in various settings show that the proposed method can steadily synthesize the separated speech of good listening quality, nearly without any interference, which is difficult to avoid in regression based methods. On the other hand, the big performance gap between the subjective and objective evaluation indicates the disadvantages of our method in the current signal-level evaluation system, which we defer to future work. We believe our proposed method can inspire another feasible avenue for exploring solutions to the cocktail party problem. 

%\section{Acknowledgement}
%Part of this work used the Extreme Science and Engineering Discovery Environment (XSEDE) ~\cite{ecss}, which is supported by National Science Foundation grant number ACI-1548562. Specifically, it used the Bridges system ~\cite{nystrom2015bridges}, which is supported by NSF award number ACI-1445606, at the Pittsburgh Supercomputing Center (PSC).

\clearpage
% \bibliographystyle{IEEEtran}
% \bibliography{mybib}
\section{References}
{
\setstretch{0.93}
\printbibliography
}

\appendix

\begin{figure*}
    \centering
    \resizebox{0.8\linewidth}{!}{
    \includegraphics[width=\linewidth]{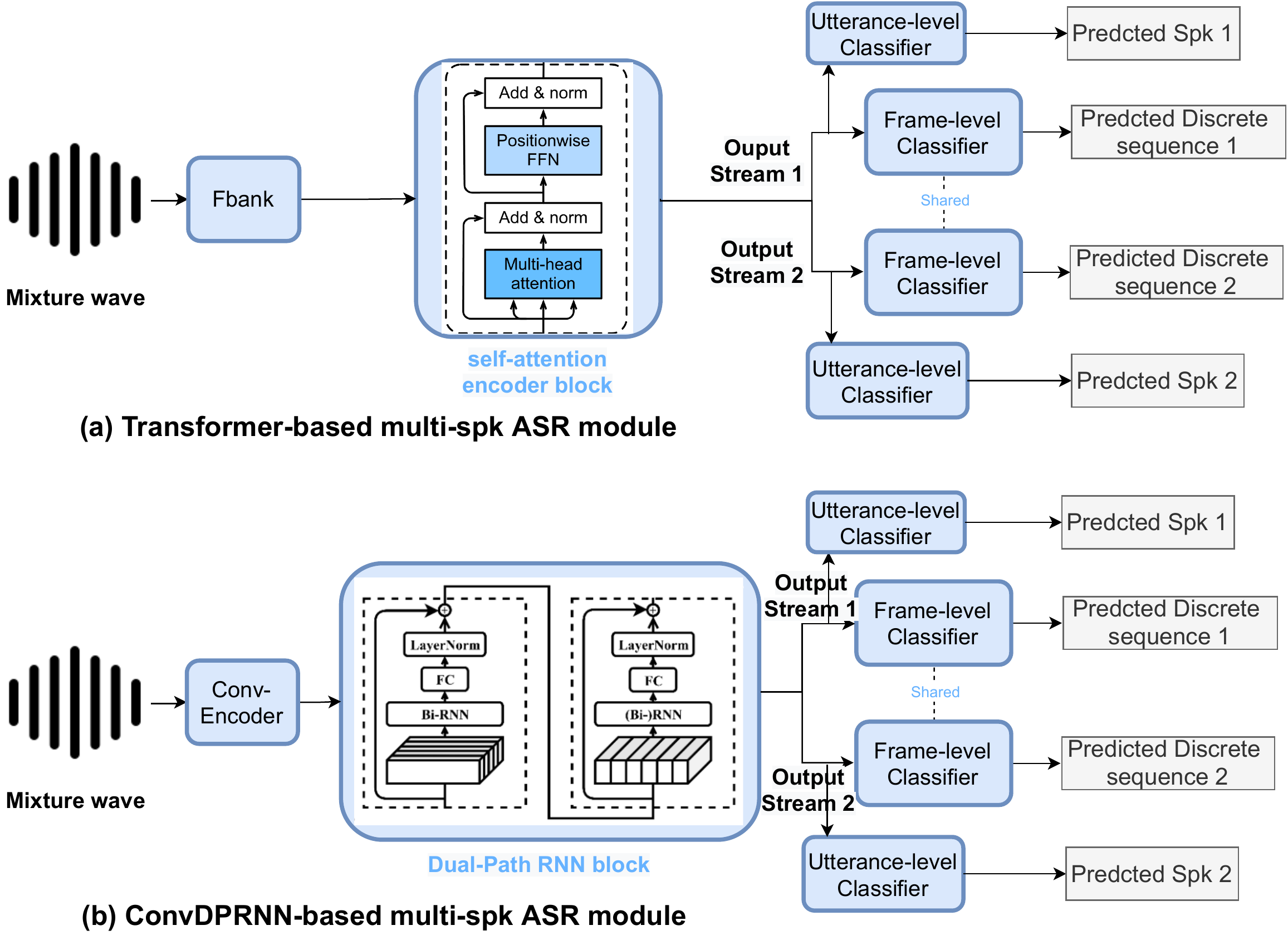}}
    \caption{Illustration of two different pseudo-ASR models we used: (a) transformer-based model, (b) ConvDPRNN-based model. Each model could be seen as a specific implementation of the pseudo-ASR module (with color of blue) in Figure~\ref{fig:model}(b). }
    \label{fig:asr}
\vspace{-0.5cm}
\end{figure*}
\section{\\The detail of pseudo-ASR module}\label{apx:asr}
As described in Sec~\ref{sssub:asr}, there are two types of ASR model we used in our experiments: (1) transformer-based models and (2) Conv-DPRNN based models, which is elaborated in the Figure~\ref{fig:asr}. We address here again that the number of output streams could be set as 1 or a number larger than 2, which should be identical with the number of speakers in given speech input (speech separation of 2 speakers or speech enhancement of 1 speaker).

In detail, for the transformer-based models, we use the similar architecture proposed by ~\cite{karita2019comparative} with FBank as the base feature extraction and multiple transformer encoder following that to output single or multiple streams. After that, for each output stream, a shared frame-level classifier is adopted to predict the discrete sequence at each frame. 
We use the transformer-based model for the experiments in speech enhancement in Sec~\ref{ssec:enh_exp}. At each block of the Transformer encoder, the network has 4 attention heads and each has dimensionalities $d^{\text{attn}}=64$. The dimensionality of the feedforward layer is $d^{\text{ff}}=2048$. A two-layer CNN with downsampling rate of 2 is used to make the time resolution consistent with the discrete symbols.

Similarly, the conventional speech separation model, e.g. Conv-TasNet, Conv-DPRNN, could be directly utilized as the feature extraction and separator part. In specific, for the results in speech separation experiments in Sec~\ref{ssec:exp_ss}, we use the 1-D convolution with 1,024 channels, 512 kernel size and the frame-shift stride of 20ms (which is consistent with the HuBERT discretization setting) to extract the base features. The same architecture of Dual-path RNN block proposed by ~\cite{luo2019dual} is used as the $separator$ with 6 layers, 256 units and 4 segment size . 

\section{\\Comparison Results for VQ-VAE}\label{apx:vq-vae}
As described before, we also implemented proposed discretization-synthesis framework with the VQ-VAE model with the MelGAN vocoder. The results compared with the baseline models and HuBERT+Hifi-GAN are shown in Table~\ref{tab:ss_vqa}. Interestingly, we find that the objective metrics with VQ-VAE + Mel-GAN are better than the HuBERT model + HiFiGAN vocoder, which is opposite with our subjective listening. In particular, there is a significant gap in the ability of speaker conversion with the VQ-VAE + MelGAN. We infer that, the lower downsampling rate (64 vs 160) with the VQ-VAE setting causes less frame shift in the generated speech $\mathbf{\hat{X}}$ compared with the ground-truth $\mathbf{X}$, which is quite important for the evaluation of used objective measures.

\begin{table*}[]
\caption{The objective evaluation in separation performance with our proposed method and some baseline models in WSJ0-2mix validation and test set.}
\label{tab:ss_vqa}
\centering
\resizebox{0.8\textwidth}{!}{%
\begin{tabular}{l|c|c|c|c|c|c}
\toprule
\multicolumn{1}{c|}{\multirow{2}{*}{\textbf{Model}}} & \multirow{2}{*}{\textbf{\begin{tabular}[c]{@{}c@{}}Discrete\\ Vocoder\end{tabular}}} & \multicolumn{5}{c}{\textbf{Separation Performance (cv/tt)}} \\ \cline{3-7} 
\multicolumn{1}{c|}{} &  & \textbf{STOI} & \textbf{SAR} & \textbf{SDR} & \textbf{SIR} & \textbf{MOSNet} \\ \toprule
(1) GT wave & x & - & - & - & - & 3.129/3.216 \\ \hline

\begin{tabular}[c]{@{}l@{}}(3) Baseline: \\ Conv-DPRNN\cite{luo2019dual}\end{tabular} & x & 0.950/0.960 & 17.41/17.09 & 16.83/16.46 & 27.43/26.90 & 2.953/3.047 \\ \hline\hline

\multirow{2}{*}{\begin{tabular}[c]{@{}l@{}}(5) Our-base\\ (Conv-DPRNN)\end{tabular}} & \begin{tabular}[c]{@{}c@{}} HuBERT + Hifi-GAN \\ Multi-spk WSJ0\textcolor{blue}{\textcircled{2}}\end{tabular} & 0.70/0.67 & -11.95/-12.44 & -12.94/-13.60 & 7.81/6.41 & 2.937/2.976 \\ \cline{2-7} 
 & \begin{tabular}[c]{@{}c@{}} VQ-VAE + MelGAN \\ Multi-spk WSJ0\textcolor{blue}{\textcircled{1}}\end{tabular} & 0.82/0.82 & 0.16/-0.95  & -0.06/-1.23  & 18.65/17.22  &2.873/2.924  \\ \hline \bottomrule
\end{tabular}%
}
\end{table*}

\section{\\Explanation of Subjective and objective metrics}\label{apx:metrics}
\subsection{Objective measures}
For the evaluation of speech separation part mentioned in Sec.\ref{ssec:exp_ss} and Tabel~\ref{tab:ss_all}, we used objective metrics, e.g., SAR, SDR, SIR, STOI, to evaluate the signal-level similarity of the predicted speech compared with the ground-truth clean signal. To be specific,
the sources-to-artifacts ratio
\begin{equation}
SAR :=10\log_{10}\frac{||s_{target}+e_{interf}+e_{noise}||^2}{||e_{artif}||^2},
\end{equation}

the source-to-distortion ratio
\begin{equation}
SDR :=10\log_{10}\frac{||s_{target}||^2}{||e_{interf}+e_{noise}+e_{artif}||^2},
\end{equation}

the source-to-interferences ratio
\begin{equation}
SIR :=10\log_{10}\frac{||s_{target}||^2}{||e_{interf}||^2},
\end{equation}
we recommend the readers to refer to the original definition proposed by Emmanuel Vincent et al. In addition, STOI (the short-time objective intelligibility) refers one objective intelligibility measure, which shows high correlation with the intelligibility of both noisy, and TF-weighted noisy speech.

For the speech enhancement related measures mentioned in Sec.\ref{ssec:enh_exp} and Tabel~\ref{tab:enh_objective}: PESQ 
, SIG, BAK and OVL, we use the provided tool from the publisher website \url{https://www.crcpress.com/downloads/K14513/ K14513_CD_Files.zip} to evaluate the results.

\begin{itemize}
\item PESQ: Perceptual evaluation of speech quality, using the wide-band version recommended in ITU-T P.862.2 (from –0.5 to 4.5).
\item SIG: the signal distortion attending only to the speech signal(from 1 to 5).
\item BAK:  the intrusiveness of background noise (from 1 to 5).
\item OVL: MOS prediction of the overall effect (from 1 to 5).
\end{itemize}

\subsection{Subjective measures}
Besides the objective measures to evaluate how much the original signal is recovered. We also use some subjective measures to test the human feeling about the quality of the given speech. In speech separation, we use the estimator (MOSNet in Table~\ref{tab:ss_all} provided by \cite{2019mosnet}, which basically is fitted to the human 5-scale mean opinion score (MOS) of the given speech. 

For the real human evaluation, we used the separation MOS defined by us in section~\ref{ssec:exp_ss}. For the noise MOS recently proposed by community, we follow the 5-scale setting that the
rater will only attend to the background. The categories of background in this sample are 5-Not Noticeable / 4-Slightly Noticeable / 3-Noticeable But Not Intrusive / 2-Somewhat Intrusive / 1-Very Intrusive.

\section{\\Procedure of the framework}\label{apx:algo}
The detail procedure of our discretization and re-synthesis could be formulated as follows:
\begin{algorithm}
    %\SetAlgoRefName{} % no count number
    \caption{The formulation of our discretization and re-synthesis framework.}
    \label{alg}
    \textbf{Input:} mixed/noisy speech  $\mathbf{M}$, oracle number of speaker  $\mathbf{K}$, ground-truth speech set $\{\mathbf{X}^k\} \in \mathbf{M}$ ;\\
    \textbf{Pre-trained:} \\
    discretization model, e.g., HuBERT, to convert $\mathbf{X}^k$ into a sequence of symbols $\mathbf{Y}^k$; \\
    Vocoder model to recover the $\mathbf{X}^k$ from the character-level sequence $\mathbf{Y^k}$; \\
    \textbf{Training:} \\
    train the pseudo-ASR model to predict all $\mathbf{K}$ target discrete sequences $\{\mathbf{Y}^k\}$ and the estimated speaker for each sequence with uPIT\cite{Kolbaek2017Multitalker_pit} strategy;\\
    
    \textbf{Test:} \\
    (1)	For the given input $\mathbf{M}$, first use the pseudo-ASR model to predict the $\mathbf{K}$ discrete sequences $\{\mathbf{\hat{Y}}^k\}$ and the estimated speaker $k$ for each;\\
    (2) Use the vocoder to synthesize the speech stream $\mathbf{\hat{X}}^k$ of each $\mathbf{\hat{Y}}^k$ on the condition of estimated speaker identity; \\ 
    (3) Evaluate the generated speech set  $\{\mathbf{\hat{X}}^k\}$ with the ground-truth $\{\mathbf{X}^k\}$
\end{algorithm}
% some others

\end{document}